# Chemical order transitions within extended interfacial segregation zones in NbMoTaW

Doruk Aksoy [1], Megan J. McCarthy [1, 2], Ian Geiger [1], Diran Apelian [1], Horst Hahn [1, 3], Enrique J. Lavernia [1], Jian Luo [4], Huolin Xin [5], Timothy J. Rupert [1, *]

[1] Department of Materials Science and Engineering, University of California, Irvine, CA 92697, USA

[2] Sandia National Laboratories, Albuquerque, NM 87123, USA

[3] Institute of Nanotechnology, Karlsruhe Institute of Technology, 76344 Eggenstein-Leopoldshafen, Germany

[4] Department of NanoEngineering, University of California San Diego, La Jolla, CA 92093, USA

[5] Department of Physics & Astronomy, University of California, Irvine, CA 92697, USA

[*] trupert@uci.edu

## ABSTRACT

Interfacial segregation and chemical short-range ordering influence the behavior of grain boundaries in complex concentrated alloys. In this study, we use atomistic modeling of a NbMoTaW refractory complex concentrated alloy to provide insight into the interplay between these two phenomena. Hybrid Monte Carlo and molecular dynamics simulations are performed on columnar grain models to identify equilibrium grain boundary structures. Our results reveal extended near-boundary segregation zones that are much larger than traditional segregation regions, which also exhibit chemical patterning that bridges the interfacial and grain interior regions. Furthermore, structural transitions pertaining to an A2-to-B2 transformation are observed within these extended segregation zones. Both grain size and temperature are found to significantly alter the widths of these regions. Analysis of chemical short-range order indicates that not all pairwise elemental interactions are affected by the presence of a grain boundary equally, as only a subset of elemental clustering types are more likely to reside near certain boundaries. The results emphasize the increased chemical complexity that is associated with near-boundary segregation zones and demonstrate the unique nature of interfacial segregation in complex concentrated alloys.

# 1  INTRODUCTION

The discovery of multi-principal element alloys has led to increased interest in microstructural engineering for these materials [1]. Also known as complex concentrated alloys (CCAs), this new domain of metallurgy presents unique opportunities to obtain seemingly contradictory material behavior such as high strength and ductility, as well as thermal and radiation damage resistance, simultaneously [2]. CCAs are typically classified as alloys with four or more elements in approximately equiatomic compositions and identified by their distinctive high configurational entropies [3, 4]. One of the extraordinary characteristics sometimes associated with CCAs is their stability against phase separation and intermetallic compound formation [2, 5]. The current lack of computational tools for predictive microstructural engineering in CCAs creates both a challenge and an opportunity in this field [6].

For high-temperature applications, refractory complex concentrated alloys (RCCAs) and high-entropy ceramics stand out among other subsets of CCAs [6]. RCCAs, such as the quaternary NbMoTaW and several quinary alloys based on this system, retain high thermal and radiation tolerances [7–9] without compromising mechanical strength [10]. For example, at temperatures greater than 1200 °C, the NbMoTaW RCCA was found to display strong resistance to strain softening [11] and superior yield strength compared to some Ni-superalloys [12]. This behavior makes NbMoTaW a strong candidate for applications for extreme conditions such as the plasma-facing parts of highly-anticipated fusion nuclear reactors [13]. Recent studies on this RCCA aim to elucidate the preferential short-range ordering behavior and its effect on strengthening mechanisms [14]. Short-range order in the NbMoTa family of alloys has also been shown to reduce vacancy migration pathways, providing a means of enhancing radiation tolerance [15].

Chemical short-range ordering (CSRO) and clustering behavior in CCAs are often linked to improved strength and increased ductility due to the ability to restrict dislocation mobility and encourage cross-slip, respectively [16–18]. Furthermore, this complex chemical landscape influences dislocation behavior even after slip initiation [14]. Energy barriers associated with defect annihilation and migration are non-uniform, unlike conventional alloys, due to the local atomic environments found in CCAs [19]. This chemical heterogeneity is also susceptible to thermal variations. There are two critical phase transitions that are believed to be significant according to specific heat capacity calculations using first-principles methods [20, 21]. Prior research places the first phase transition around room temperature, where maximum specific heat capacity is observed due to Nb and W segregation that results in partial breaking of predominant Mo-Ta B2 structures [16, 22]. The second transition, which originates from a chemical order-disorder reaction, is predicted to be in the 700-1300 K range [16, 20]. There are many material and thermodynamic properties that influence the chemical ordering behavior including electronegativity [23], misfit volume [24], enthalpies of mixing [25], chemical affinities [26], and particle size differences [27], all of which are interlinked with enthalpy of mixing being a major factor that have chemical and/or strain origins.

The energy landscape of a CCA gets more convoluted when crystalline defects are considered. Specifically, grain boundaries require special attention due to their ability to directly alter the mechanical behavior of materials. For example, certain structural motifs associated with grain boundaries are more inclined to nucleate dislocations [28–31]. Thus, it is imperative to consider the role of grain boundary structure pertaining to segregation and interfacial response [32–36]. For example, the stability of grain boundaries in multi-component systems depends on the spectrum of segregation energies associated with that boundary [37]. In a grain boundary network, segregation behavior is not only connected to available sites in a boundary, but also to other boundaries in the network, due to the energetic competition between sites, as well as solute-solute interactions [38–40]. Interfaces in CCAs differ from conventional alloys because they can host chemically complex environments, which can be utilized to stabilize high-temperature nano-grained alloys [41], or hinder boundary migration through grain boundary roughening mechanisms [42]. Furthermore, solute segregation to grain boundaries can promote local ordering and results in solid solution hardening [16], as well as interfacial disordering. The local lattice distortion effect of CCAs [27, 43] can be exacerbated due to excess volume availability in grain boundaries. There are multiple studies that address the CSRO and clustering behavior within the first and the second nearest-neighboring shells in the NbMoTaW refractory CCA [14, 44–47]. However, most of these studies treat the innate chemical complexity of the system in the bulk phase only, without considering the role of defect structures such as interfaces. One of the ways to approximate the effect of chemical complexity in the system is to develop an average-atom interatomic potential [48]. In this method, all atoms in the model are replaced by an artificial atom species that encapsulates the mechanical and chemical properties of the CCA that it was developed for. For instance, Farkas utilized this model to investigate grain boundary structure in an FCC CCA [49]. These examples demonstrate the significance of local atomic environments that are associated with interfaces, and their role in the complex relationship between interface structure and CSRO tendencies. To provide insight into this relationship, McCarthy et al. [50] studied faceted $\sum 11$ tilt boundaries of two equiatomic face-centered cubic (FCC) CCAs (CrFeCoNiCu and CrFeCoNi), and documented co-segregation regions near grain boundaries that diverged from the traditional definition of the grain boundary. Their investigation of the segregation behavior in these boundaries revealed, in addition to defected or non-FCC atoms, the presence of structurally distinct local atomic environments characterized by reduced atomic volumes. The widths of the grain boundaries before segregation were between 5 and 10 Å; after segregation, the widths of these extended segregation zones were in the range of ∼11 and 17 Å. This finding suggests that the grain boundary-affected region spans further than just the defected atoms, and that these zones can be especially important in CCAs where a number of elements are available for chemical partitioning.

In this work, the role of the interface structure on compositional variance, CSRO, and clustering behavior in the refractory CCA NbMoTaW is investigated. In contrast to the alloy family from the prior study by McCarthy et al. [50], the NbMoTaW system exhibits a strong tendency for CSRO, allowing for an understanding of the competition/synergy between structural and chemical aspects of boundary segregation. *Near-boundary segregation zones* (NBSZs) are found adjacent to the grain boundary and triple junction regions and exhibit transitory behavior connecting the interfaces to the bulk regions. The effect of CSRO is apparent on the interfacial atomic fractions of constituent elements. For example, Mo and Ta atomic fractions are closely linked, which points to a B2 structural pattern. In fact, exploring the A2-to-B2 transition within NBSZs reveals a common pattern across all models. As one moves away from the boundary, the predominant structure of the interface (Nb-Nb A2) yields to the B2 structure observed heavily in bulk. Although NBSZs were also identified in the FCC-structured CCA in the prior study by McCarthy et al. [50], no structural transition from the interfaces to the bulk regions was observed. This transition point is intimately related to the chemical order associated with this alloy, and is affected by not only the interfacial structure but also the grain size and temperature. In the analysis on all pairwise elemental interactions, some interactions such as the Nb-Ta pair-interaction do not appear to be affected by the presence of an interface. In contrast, there are other elemental pairs that have preferential chemical order depending on their proximity to interfaces, such as the Nb-Nb and Mo-Ta pairs that exhibit antagonistic short-range order behavior. Finally, a more detailed investigation that considers individual grain boundary and triple junction regions suggests that some regions are more inclined to experience atomic clustering. These findings provide a thorough understanding of NBSZs in systems with strong CSRO and will facilitate the design of interfacial properties of future RCCAs.

## 2    METHODS

In this work, atomistic simulations are performed using the Large-scale Atomic/Molecular Massively Parallel Simulator (LAMMPS) [51] software package. Atomic visualization is accomplished with OVITO [52] and the remaining data visualization uses Python libraries Matplotlib [53] and Seaborn [54]. The interatomic potential used in this work [17] is a machine learning interatomic potential that was developed in accordance with the Moment Tensor Potential [55] methodology. This methodology utilizes invariant vectors to represent local atomic environments. During the fitting procedure, the results of each new iteration of the potential is compared to results obtained from density functional theory simulations. Material properties such as the melting temperature, unstable stacking fault energy, and elastic constants, as well as chemical short-range order behavior and dislocation mobility are in good agreement with experiments and density functional theory results [17].

NbMoTaW thin film models are prepared with Atomsk [56], an open-source program that allows for the creation of atomic-scale models with pre-defined attributes. To model four distinct hexagonal shaped grains, seeds are placed at the center of each grain. These four grains are separated by tilt grain boundaries with rotation angles of approximately 79.3°, -49.7°, -30.4° and

22.2° around the (001) axis, corresponding to Σ13 [001]/[5$\bar{1}$0], Σ25 [001]/[3$\bar{4}$0], Σ53 [001]/ [9$\bar{5}$0], and Σ97 [001]/[4$\bar{9}$0], respectively. The boundaries of the model are fully periodic in all directions, to reflect the columnar grained structure often observed in deposited NbMoTaW thin films [8]. The average grain size, $d$, of NbMoTaW thin films deposited using magnetron sputtering on a (100) Si substrate at the room temperature is within the range of 5-20 nm [15, 57]. Accordingly, three different grain sizes (approximately 5, 10 and 20 nm) are selected, with a slab thickness of 6.5 nm in the Z-direction of Figure 1(a). The relative volumetric differences in the individual grains are less than 0.8%. Due to the similar bond lengths between the four constituent elements (Nb, Mo, Ta, and W), pure Nb models are first created. Next, these models are converted into random solid solution CCAs by randomly assigning chemical types to atomic sites corresponding to an equiatomic composition without any impurities (i.e., an atomic fraction of 0.25 for each element). To remove any residual strain created by changing the chemical composition, overlapping atoms within a cutoff distance of 1.9 Å are deleted while allowing the simulation box to relax during the energy minimization procedure. The model is then annealed for 100 ps with an isobaric, isothermal (NPT) ensemble using an integration timestep of 1 fs, followed by relaxation using a conjugate-gradient energy minimization procedure. Overall, 12 different models were simulated with varying grain sizes at 300 K and a constant grain size of 10 nm at temperatures ranging from 100 K to 1900 K.

To obtain the final equilibrate configuration with segregation to the defect regions, hybrid Monte Carlo and molecular dynamics (MC/MD) simulations were performed. During these simulations, for every other MD step (integration time step of 5 fs), 1% of all atoms were swapped between every pair combination of the constituent elements in the CCA according to a Metropolis MC algorithm. The equilibrium state of each simulation is assessed by tracking the gradient of potential energy and of interface atomic fractions. The gradients of potential energy for each model converges to a value less than 0.01 meV at the conclusion of the MC/MD simulations. The grain boundary concentrations stabilize far before that, at approximately 8 ps. Additional discussion on configurational stability can be found in Supplementary Note 1. Fig. 1(a) shows the grain structure of a 20 nm grain size model where the interface and bulk sites, identified with adaptive common neighbor analysis method in OVITO [58], are denoted by light gray and blue, respectively. Fig. 1(b) shows the equiatomic random solid solution state of this model, with the colors corresponding to the constituent elements Nb, Mo, Ta, and W being cyan, magenta, yellow and black, respectively. Fig. 1(c) shows the equilibrated structure corresponding to the final snapshot of the hybrid MC/MD simulation at 300 K.

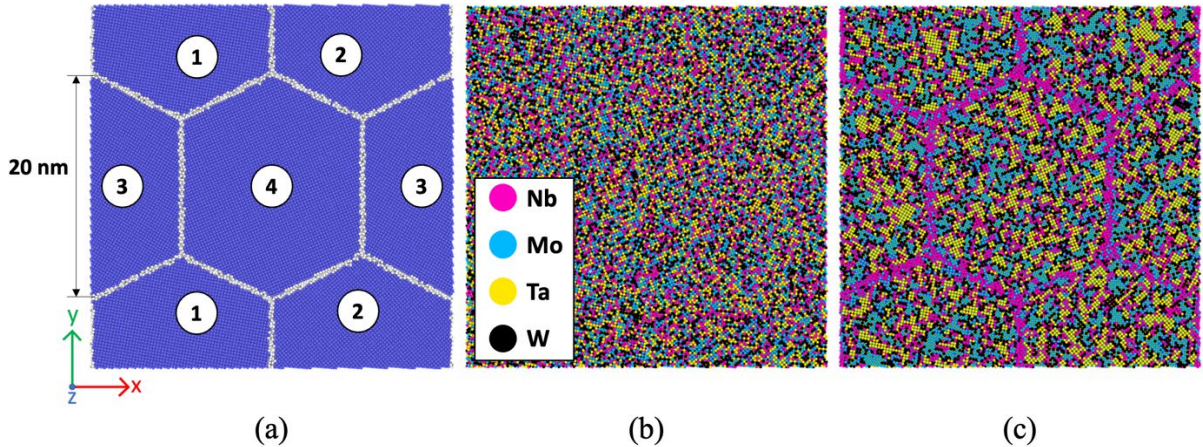

(a)                    (b)                    (c)

**FIG. 1. 20 nm grain size simulation models showing (a) the starting grain structure with interface (light gray) and bulk (blue) sites identified with adaptive common neighbor analysis method, (b) the random solid solution state with equiatomic composition, and (c) the equilibrated structure corresponding to the final snapshot of the hybrid MC/MD simulation at 300 K. Four distinct grains are found in the model, as illustrated in Fig. 1(a).**

Local variations in the chemical composition near the interface were studied to better understand the role of the interface structure on segregation behavior after the hybrid MC/MD simulations. These regions are shown in Fig. 2(a), illustrated with a gradient color scheme drawn on top of interfacial and near-interfacial regions. The numbers on the color bar represent radial increments, with "0" representing the aggregate of all defected interface sites, as identified by adaptive common neighbor analysis. Moving further from the aggregate of all interface sites, 1 Å thick slices are taken by expanding the region radially. These slices are utilized in the investigation of chemical composition and CSRO behavior.

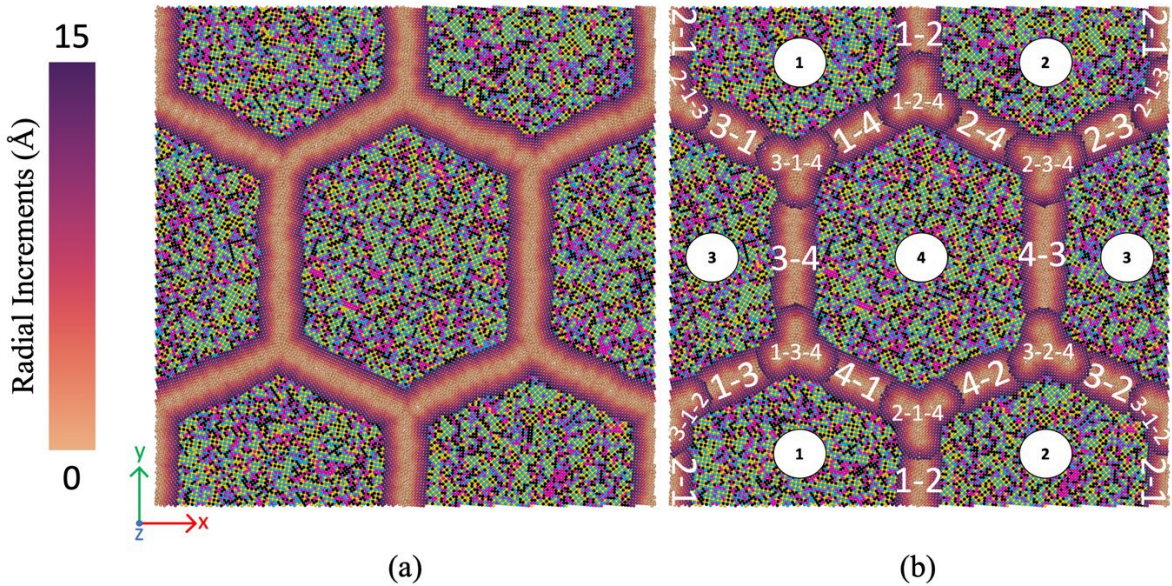

FIG. 2. (a) Radial expansion spanning from interface sites to bulk allows for spatial variations in local chemical composition to be quantified. Radial increments illustrated with the color bar increase from lighter to darker colors as one moves away from the interface, with all defect atoms sites are collected within the "0" radial increment. (b) Illustration of the naming convention for specific grains and their associated grain boundary and triple junction regions, shown for the 20 nm grain size model at 300 K.

In addition to separating and analyzing the interface region as a whole, the segregation behavior in individual grain boundary and triple junction regions can be extracted. 20 distinct regions (12 grain boundaries and 8 triple junctions) can be identified and analyzed in this model. The first step of our partitioning procedure is to identify atoms at the center of triple junction regions, where three boundaries meet. Then, the centers of the grain boundary regions are denoted as the mid-distance between two triple junction regions. The structure between the adjacent triple junction and grain boundary regions can be locally distinct and thus exhibit unique segregation patterns [40]. To identify where these structural transitions take place along a grain boundary as it approaches a triple junction, a preliminary analysis is performed by calculating mean atomic volumes and atomic concentrations in a grain boundary plane, which is shown in Supplementary Note 2. The results suggest that a grain boundary is well described by the middle 75% of its length, which stretches from the center of one triple junction to another. Beyond this point, structural differences are attributed to triple junctions and should be treated separately. A step-by-step guide for partitioning and slicing an individual grain boundary and triple junction regions is provided in Supplementary Note 3. This partitioning of grain boundary and triple junction sites is shown in Fig. 2(b), overlaid with the naming convention. For example, the triple junction region where grains 2, 3, and 4 meet is denoted as $TJ^{2-3-4}$, similarly the grain boundary between grains 1 and 4 is denoted as $GB^{1-4}$.

Finally, CSRO and clustering behavior is also analyzed within and near the grain boundary region. One of the most common ways to quantify CSRO is through Warren-Cowley parameters [59], which expresses whether the probability of occupation for neighboring atoms in a shell of a central atom is higher or lower compared to its concentration in the system. In this work, a modified version of the Warren-Cowley parameters is utilized based on Ref. [45], which is advantageous because this version ensures that a positive Warren-Cowley parameter is of the attractive nature regardless of the interaction type (i.e., same-specie or inter-species). In this method, the CSRO is calculated as:

$$\alpha_{ij}^k = \frac{P_{ij}^k - c_j}{\delta_{ij} + (-1)^{\delta_{ij}} c_j} \quad , \tag{1}$$

where $P_{ij}^k$ is the probability of occupation by a $j$ atom within $k$th nearest-neighbor shell when atom $i$ is at the origin, $c_j$ is the concentration of atom $j$ in the CCA, and $\delta_{ij}$ is the Kronecker delta function. In this work, pairwise interactions are presented as one of two categories according to their constituent element placement in the periodic table: (1) inter-species (bonding between different species, e.g., Nb-W) and (2) intra-species (bonding between the same species, e.g., Nb-Nb). Identification of the phase transitions occurring at critical temperatures that were previously mentioned in Section 1 requires an in-depth exploration of clustering behavior and local crystal lattice structures. In accordance with the study by Körmann et al. [20], the clustering behavior around room temperature was investigated for the NbMoTaW CCA using the A2 (W) (Strukturbericht designation) and B2 (CsCl) structures with Polyhedral Template Matching [51] in OVITO. Only the first-nearest neighbors are used for the analysis of A2 and B2 structures.

# 3  RESULTS AND DISCUSSIONS

## 3.1  Segregation Behavior at and near the Interfaces

The chemical concentration at interface sites (i.e., those which are identified as defects) obtained after hybrid MC/MD simulations at 300 K is shown in Fig. 3 for the three grain sizes considered in this work. All three models start as random solid solutions with equiatomic distribution and show the development of similar grain boundary segregation. The interfaces display a Nb-rich composition, while Ta and W are depleted from interface sites. The slight increase and subsequent decrease of the Mo fraction can be explained by its predisposition for the formation of Mo-Ta B2 clusters. There is a temporary increase when there is significant Ta content at the interface region, followed by migration of any B2 structure to the bulk once Ta depletion occurs.

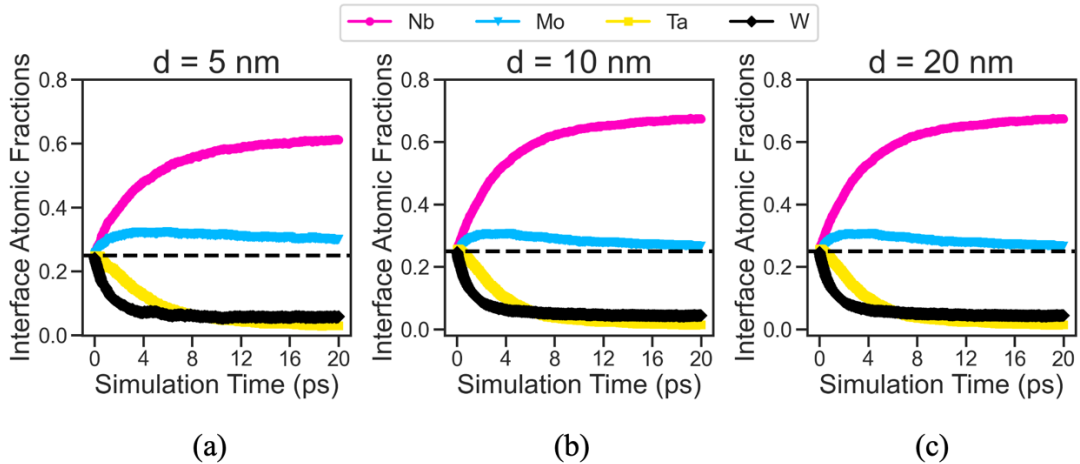

**FIG. 3.** The atomic fractions of constituent elements at the interfaces (defect atoms only) during the hybrid MC/MD simulations for (a) 5 nm, (b) 10 nm, and (c) 20 nm grain size models at 300 K.

As evident from Fig. 3, changing the grain size within this range does not appear to affect the segregation behavior at the interfaces drastically. Nb is the primary segregant in all cases, although the interfacial Nb composition increases slightly with increasing grain size. This can be explained by the fact that interfacial fraction, or the fraction of the overall simulation cell comprised of the defected interfacial regions, increases dramatically as grain size is reduced (15.9% and 4.1% for the 5 nm and 20 nm grain size samples), meaning the smaller grain sizes may have more potential segregation sites than Nb atoms available to fill them. The decrease in grain boundary Nb concentration between the 5 nm and 10 nm grain sizes (61.2% vs. 67.4%, respectively) leads to some increase in the interfacial Mo content, suggesting that Mo can fill some of the potential segregation sites when there is not enough of the stronger segregating species.

An understanding of the local composition nearby a given boundary is also needed for a full picture of the segregation state. In Fig. 4, compositional variation within each slice from the interface to the bulk is shown for the three models (5, 10, and 20 nm grain size) at 300 K. For each of the models, three zones can be identified which have distinct atomic volume regimes, in addition to different segregation behavior corresponding to interface, bulk, and transitory zones. Interface in this context is defined as the collection of grain boundary and triple junction defect sites. Relatedly, we define the transitory zones as NBSZs, bound by the interface region on one side and the point at which segregation behavior converges to bulk behavior on the other. NBSZs are denoted in Fig. 4 by the shaded semi-transparent boxes. NBSZs indicate that the change in segregation behavior due to the presence of an interface is not limited to the defected atoms, but also extends into the crystalline region. This behavior is also accompanied by variations in local atomic volume within these NBSZs, which can be seen in Fig. 4 (b), (d), and (f). At the interface, Nb, Mo, and Ta occupy sites with higher atomic excess volume than their bulk counterparts. In contrast, W occupies lower atomic excess volume sites compared to the bulk. The NBSZs serve as transition regions where the per-species atomic volumes revert to the bulk value over an extended region. The findings shown here generally agree with the discovery of NBSZs in FCC CCAs by McCarthy et al. [50], although the exact trends are different. In all three models, the compositional imbalance between Nb, Mo, and Ta at the interface region reaches an equilibrium point at the equiatomic composition, and then eventually transitions to the bulk composition that defines the grain interior. This point represents a *structural transitioning point (STP)*, where the dominant structure at the interface (Nb-Nb A2) is overtaken by the dominant structure at the bulk (Mo-Ta B2). Existence of B2 structure in BCC materials is well documented in the literature [14, 16, 21, 22, 44–47, 60, 61]; however, most existing studies have focused on behavior in the bulk phase. Further investigation is needed from other CCAs to confirm whether STPs are *de facto* indications of structural transitions between interface and bulk regions, but these results suggest that, at a minimum, there must at least be two different dominant CSRO structures between the interface and the bulk for this point to be observed. Therefore, a material without CSRO such as the FCC alloys studied by McCarthy et al. [50] would not exhibit STPs. Here, the segregation behavior in Fig. 4 shows that as the grain size is decreased, the structural transition requires fewer and fewer atomic layers (i.e., smaller distance from the defective atoms). One of the factors that could contribute to this behavior is the depletion of segregating species from the bulk regions in small grain size models. For example, as grain size is decreased, the Nb concentration in the grain interior decreases. The increasing grain boundary volume fraction with reduced grain size means that more Nb atoms are removed from the grain interior. The segregation patterns of different grain sizes shown in Fig. 3 show evidence of this effect as well, as Mo begins to segregate more as grain sizes decreases, because there is no longer enough Nb to fully decorate the grain boundaries.

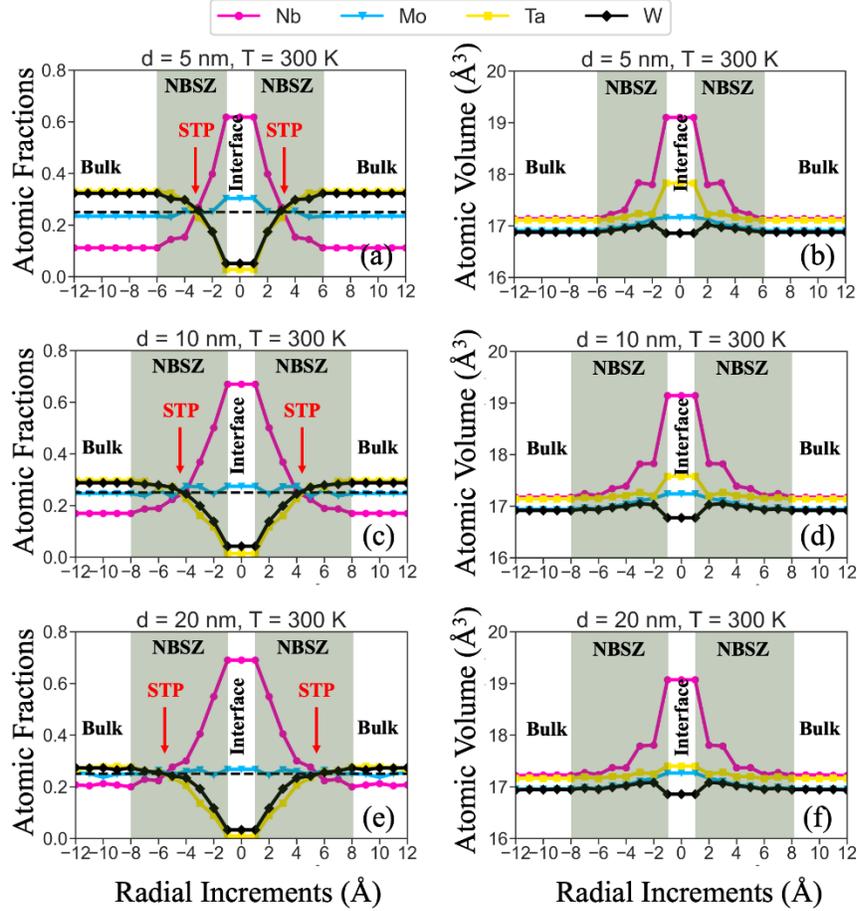

**FIG. 4. Compositional variation as a function of radial expansion originating from the interfaces for (a) 5 nm, (c) 10 nm, and (e) 20 nm grain size models at 300 K. The change in atomic volume with radial expansion originating from interfaces for (b) 5 nm, (d) 10 nm, and (f) 20 nm grain size models at 300 K. Gray boxes separate NBSZs from sites exhibiting interface or bulk behavior, and structural transition points (STPs) are shown by red arrows.**

The NBSZs and STPs defined above are clear examples of extended structural and chemical patterning in the vicinity of the interface. Chemical clustering and segregation are known to be strongly dependent on temperature, so as a next step, the evolution of these features with varying temperature is studied. Local chemical concentration in the 10 nm grain size model at temperatures ranging from 100 K to 1900 K is shown in Fig. 5, with the local atomic volume for the same models shown in Fig. 6. Following the same methodology used in Fig. 4, the sizes of the NBSZs are obtained by investigating the atomic volume change shown in Fig. 6. Similarly, STPs are identified as the point where Mo and Ta interfacial atomic fractions overtake Nb. According to Fig. 5, the structural transition back to bulk behavior requires fewer atomic layers as the temperature is increased up to 1100 K. The role of the lattice expansion effect due to temperature increase is not responsible for this behavior, since a ~20% decrease in NBSZ size is accompanied by ~3% increase in atomic volumes (i.e., lattice expansion would increase the apparent NBSZ size, here the opposite is found). Fig. 6 shows that higher temperatures are associated with higher atomic volumes within the grain boundaries in general. Analyzing the elements separately by their atomic volumes indicates that W and Ta atoms at the interface have lower volumes compared to NBSZ and bulk regions at lower temperatures. This behavior changes with increasing temperature though, with all elemental atomic volumes in the interface region above their bulk and NBSZ counterparts at higher temperatures. The increase in interface atomic volumes stabilizes around 1100 K, which is also reflected in the NBSZ widths, suggesting reduced chemical effects above that temperature.

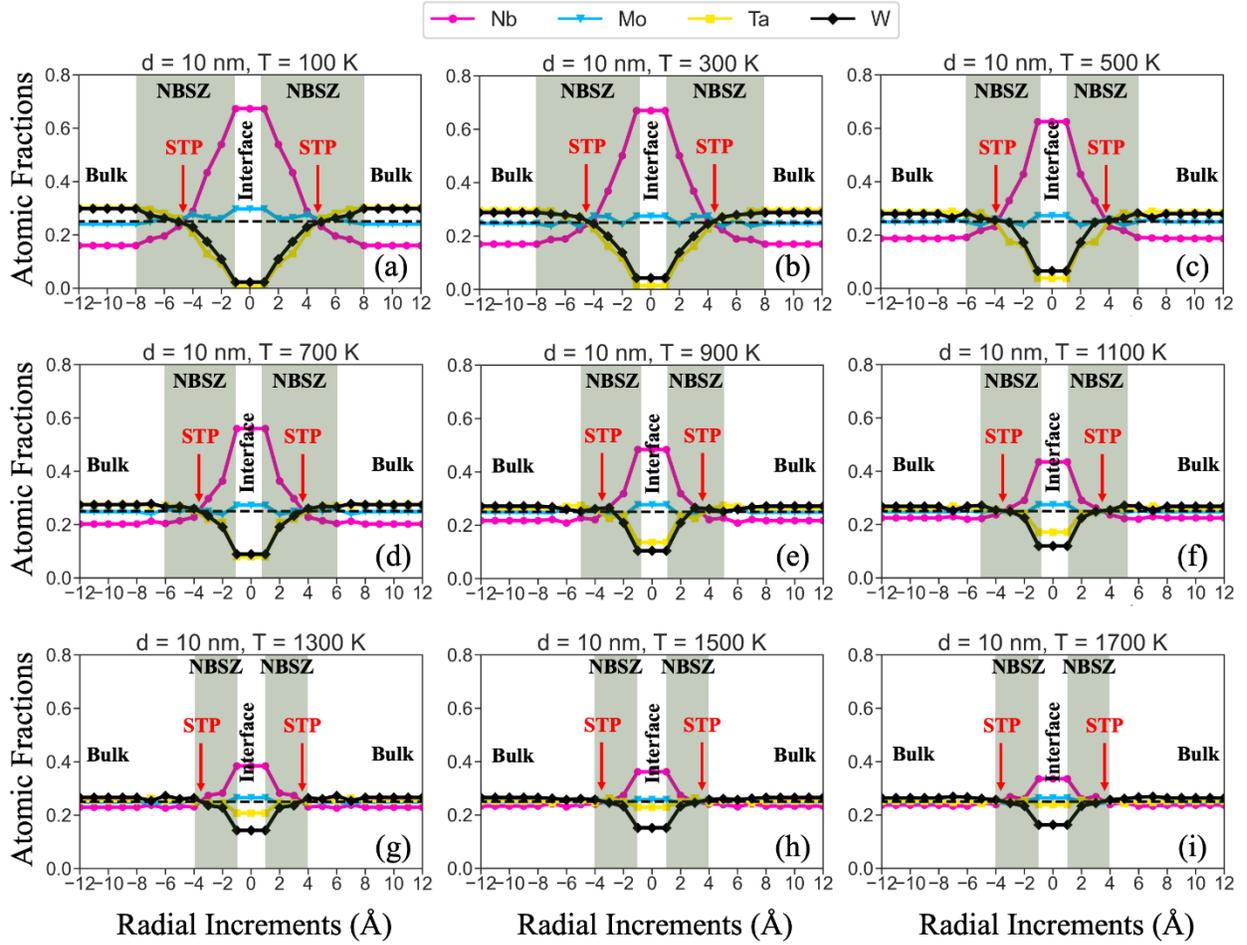

FIG. 5. Segregation behavior near the grain boundaries for 10 nm grain size models at temperatures ranging from 100 to 1700 K.

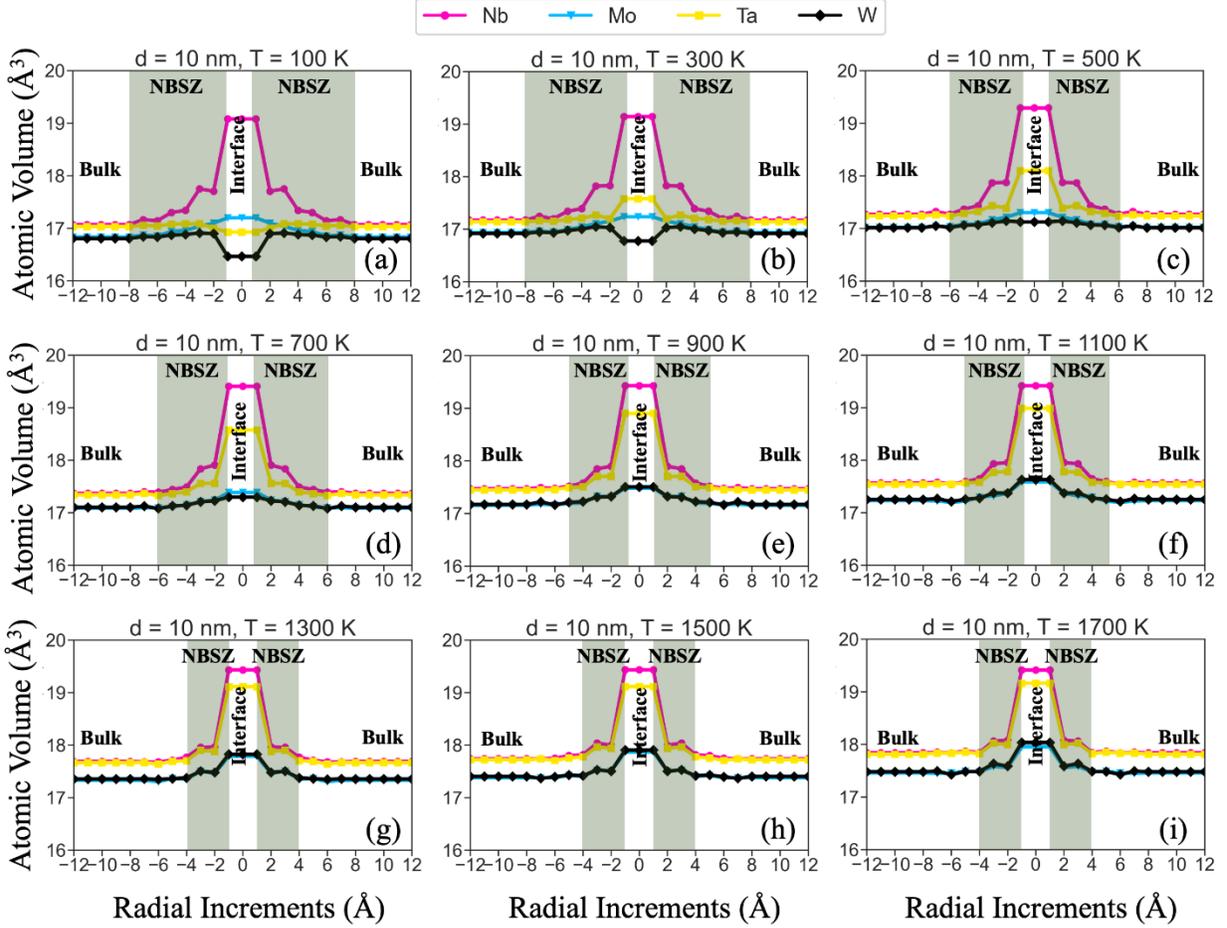

**FIG. 6. Atomic volumes near boundaries for 10 nm grain size model at temperatures ranging from 100 to 1700 K.**

Fig. 7(a) shows the widths of the interface, NBSZ, and STP regions as a function of temperature. We note that the interface width is taken as the average (~7 Å) of all 10 nm grain size models at different temperatures, to reduce subtle variations in width due to thermal fluctuations and allow for easy comparison of the NBSZ and STP regions, which are the focus of this study. As a whole, these results show that the NBSZs, where the presence of an interface alters the segregation landscape beyond defected atoms, extend the actual interface-affected region to 2-3 times as large as the interface width alone. The width of this altered region decreases as temperature increases, with a plateau observed above ~1100 K. Fig. 7(b) compiles the NBSZ, STP, and actual interfacial widths for different grain sizes at 300 K. It is important to put the increased width of the interface-affected zone (NBSZ plus interface) into proper context. From the data shown in Fig. 7, roughly 2/3 of the segregation landscape is not accounted for in an analysis that only considers traditional interface sites. This effect can be very substantial in nanocrystalline metals, where the grain boundaries make up a large volume fraction of material. For example, using the calculations of Palumbo et al. [62], in which random, equiaxed grains are modeled as tetrakaidecahedra, a material with 20 nm grain size and boundary width of 0.7 nm would have a grain boundary volume fraction of ~10%. In contrast, the same 20 nm grain size with a 2.3 nm boundary width (accounting for both defected atoms and the NBSZ) would have a grain boundary volume fraction of ~31%. Hence, the existence of NBSZs will make the grain boundary region even more influential in nanocrystalline CCAs than it is in pure metals or simpler alloys.

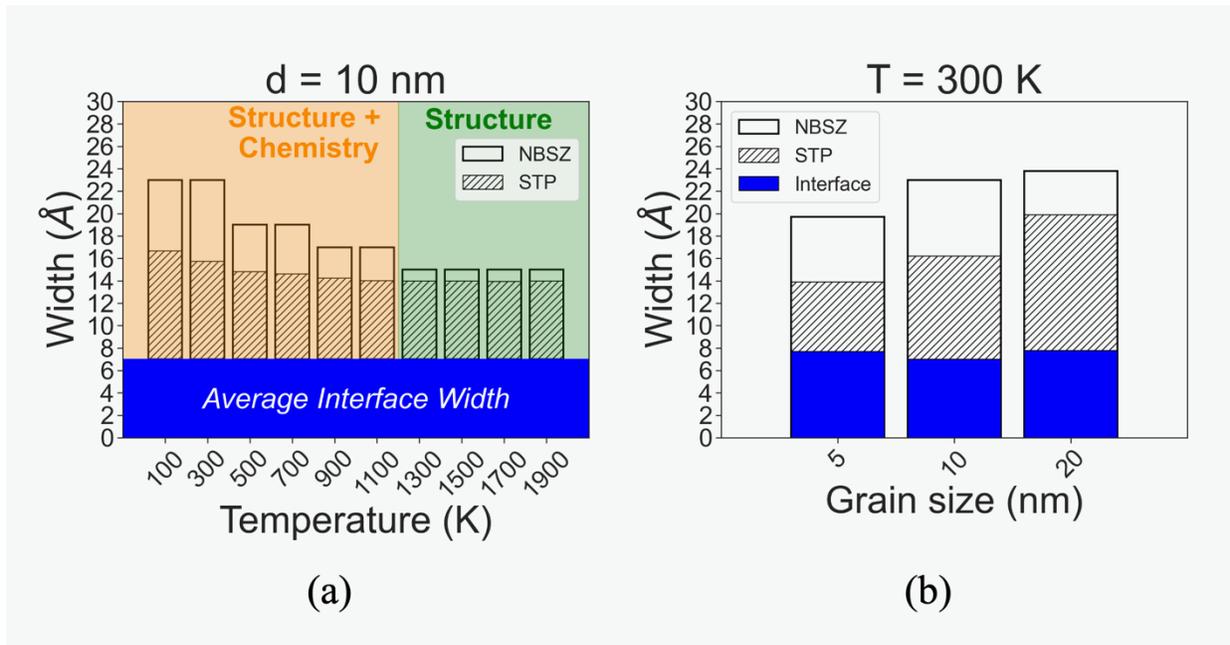

(a)                    (b)

**FIG. 7. (a) NBSZ widths for a 10 nm grain size model at temperatures ranging from 100 to 1900 K, with the average interface width marked as well as the STP location. (b) Measurements of the interface-affected zone for different grain sizes at 300 K.**

## 3.2 Chemical Short-Range Ordering and Clustering Behavior

The increased variability in microstructures observed in CCAs can amplify or reduce the effect of solute-solute and solvent-solute interactions on interfacial segregation. In this work, we identify the role of interfacial structure on the CSRO behavior by separately analyzing bulk and interface sites. Then, we examine the chemical ordering in individual segments of the interfacial region and their corresponding NBSZs.

Fig. 8 shows the effect of temperature on inter- and intra-species short-range order at interface and bulk for the 10 nm grain size model as a function of temperature. Looking at temperatures above 1100 K, a general location can be observed after which CSRO is very small and does not vary with temperature, marking the order-disorder transition. This temperature value is consistent with prior literature [45], and its effect can also be seen in Fig. 7(a), where the STP width is invariant to further temperature increase. As this temperature aligns with the disappearance of CSRO in this alloy, the extended segregation regions above this temperature originate from primarily structural effects. In contrast, at 1100 K and below, complementary structural and chemical effects are active and lead to thicker NBSZs. Another observation that agrees with previous studies is that the Mo-Ta pair exhibits the strongest attractive interaction, followed by Ta-W and Nb-Mo pairs [45]. Different from the traditional methodology, which involves investigating the CSRO on CCAs by considering all atomic sites, here we separate the short-range order behavior into separate bulk and interface data sets. By doing so, strong Mo-Ta interaction observed in bulk sites is found to be suppressed by weaker Mo-Ta interaction at the interfaces. Interestingly, the repulsive Ta-W interaction at interfacial sites changes to attractive when this pair is observed in bulk sites. This behavior is also observed during the investigation of atomic volume change with temperature, in Figs. 4 and 5. With increased temperature, the atomic fractions of Ta and W in the interface region are increased. As this occurs, the atomic volume of both of the elements increases due to the repulsive interactions between them. In contrast, at lower temperatures, the interfacial atomic fractions of Ta and W are much lower, which means very few Ta-W pairs are found and their individual atomic volumes are smaller. In the case of same-element pairs, the Nb-Nb interaction is predominantly attractive. In contrast to the previous study on NBSZs in the faceted CrFeCoNi CCA boundaries [50], the bulk region in this work exhibits strong CSRO tendencies. In addition to pair-interaction thermodynamics, this could be due to the differences in crystalline structures, as previous research suggests FCC CCAs having weaker CSRO at low temperatures compared to the strong CSRO shown in Fig. 8 [63–65]. Accordingly, the CSRO differences between FCC and BCC ordered structures could be a contributing factor to the increased strength-ductility synergy in FCC CCAs. According to this theory, clusters that favor FCC structure tend to increase the ultimate strength of the material, as opposed to clusters favoring BCC structure, which was hypothesized to increase ductility [66].

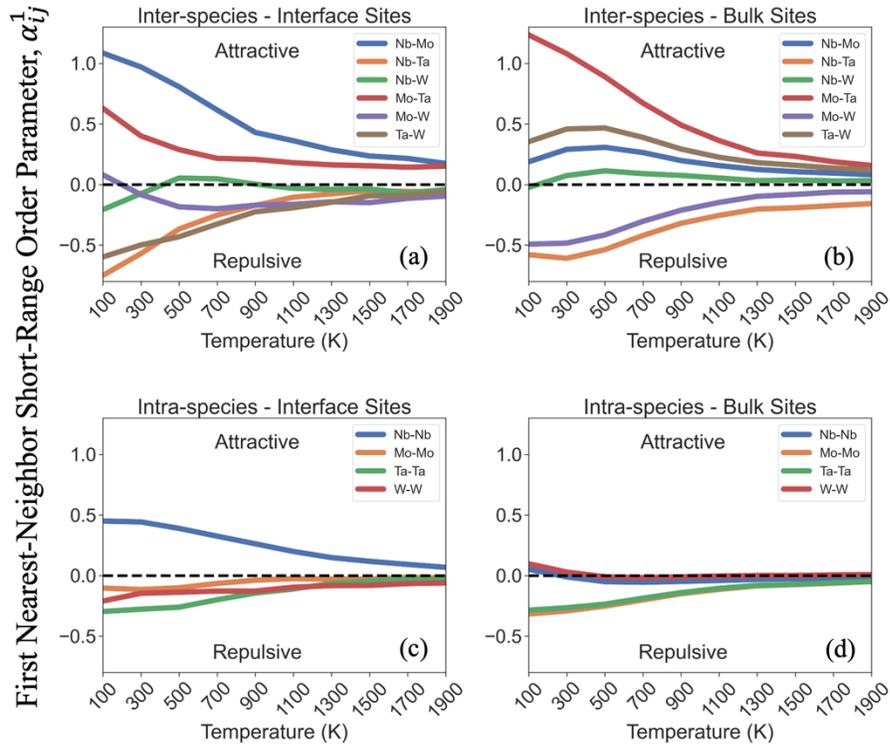

**FIG. 8.** Inter-species chemical short-range ordering behavior for (a) interface, and (b) bulk sites. Intra-species chemical short-range ordering behavior for (c) interface, and (d) bulk sites. Note that a positive ordering parameter indicates an attractive behavior for pair interactions.

The effect of CSRO on the NBSZ and STP of the entire interfacial region is next investigated. The short-range order parameters for each elemental pair are shown in Fig. 9, where the values are averaged over each slice in the entire interface region of the 20 nm grain size model at 300 K. This analysis shows that the CSRO behavior stabilizes inside the NBSZs, at the STP, after which it returns to the bulk-like behavior. Provided that the clustering behavior is observed in the simulation (i.e., at temperatures below the order-disorder transition), reducing the grain size or increasing the temperature result in convergence to bulk behavior in fewer atomic layers. While some pair interactions (e.g., Nb-Ta) have similar chemical order behavior regardless of whether the atoms are at an interface or in the bulk region, other interactions (e.g., Nb-Mo) exhibit transient behavior. Of the interactions that are affected by the presence of an interface, Nb-Nb and Mo-Ta interactions can be utilized as surrogates for the A2 and B2 structures, respectively, as these atomic pairings provide nearly all of the clusters with these order types. We note that Mo-Ta B2 pair clusters can have either Mo or Ta as the central atom in a B2 structure.

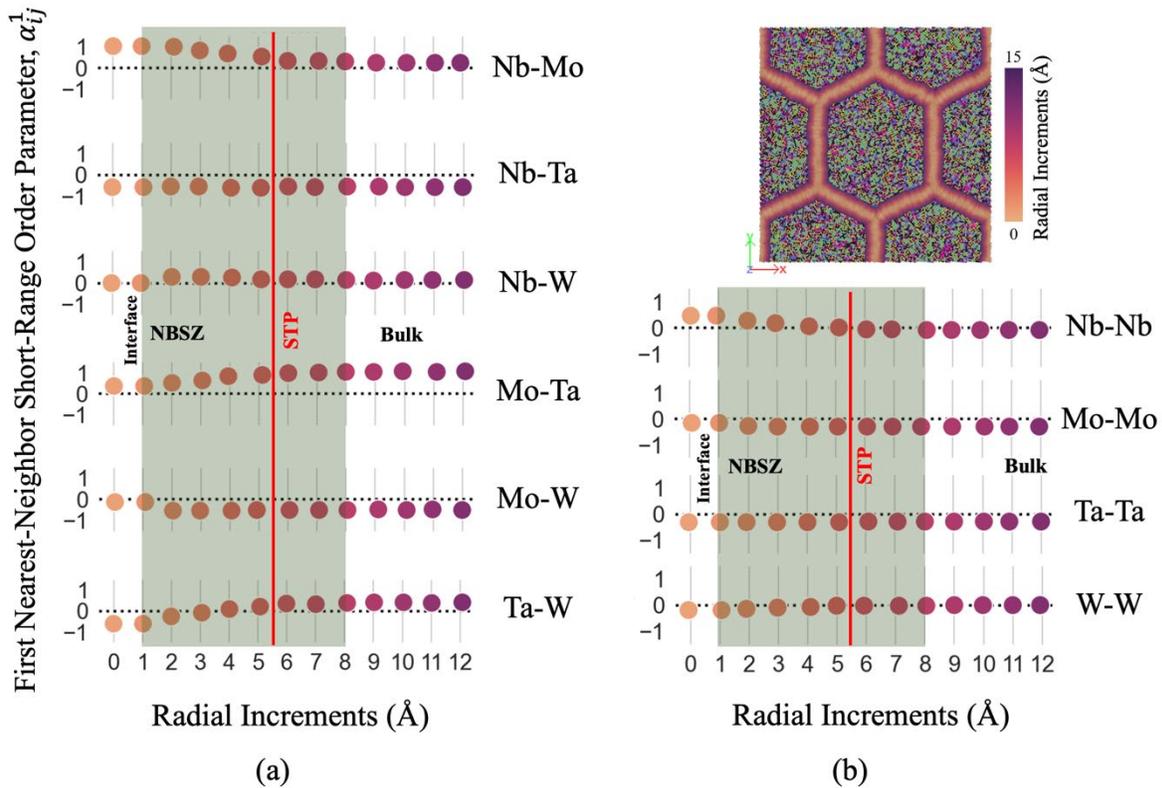

**FIG. 9. CSRO parameters near boundaries for each elemental pair in the 20 nm grain size model at 300 K for (a) inter-species and (b) intra-species pairs.**

The segregation trends between Nb-Nb and Mo-Ta clusters indicate an antagonistic relationship. The Nb-Nb pair transitions from ordered to disordered with increased radial distance from interfaces, whereas the Mo-Ta pair transitions from disordered to ordered. This transition can be clearly observed in the radial analysis of the interface and near-interface region in Fig. 10(a), which shows the 20 nm grain size model at 300 K where Nb-Nb clusters with A2 structures (pink) are observed predominantly at the interface region and Mo-Ta B2 structures (yellow and blue clusters) are found in the NBSZs. To get a better idea of the clustering mechanisms associated with individual grain boundaries and triple junctions, the cumulative number of Mo-Ta B2 pairs obtained by radial expansion is plotted in Fig. 10(b), with the number of clusters within a specific interface shown as different colored stacks. Focusing on the vicinity of individual regions, the most and the least cumulative number of Mo-Ta B2 clusters are $TJ^{3-2-4}$ with 42 clusters and $TJ^{1-2-4}$, $TJ^{1-3-4}$, and $GB^{2-4}$ with 7 clusters each. This result shows that there can be high disparity among individual interfaces in terms of number of vicinal Mo-Ta B2 clusters, with this variation likely arising from differences in the interfacial structure.

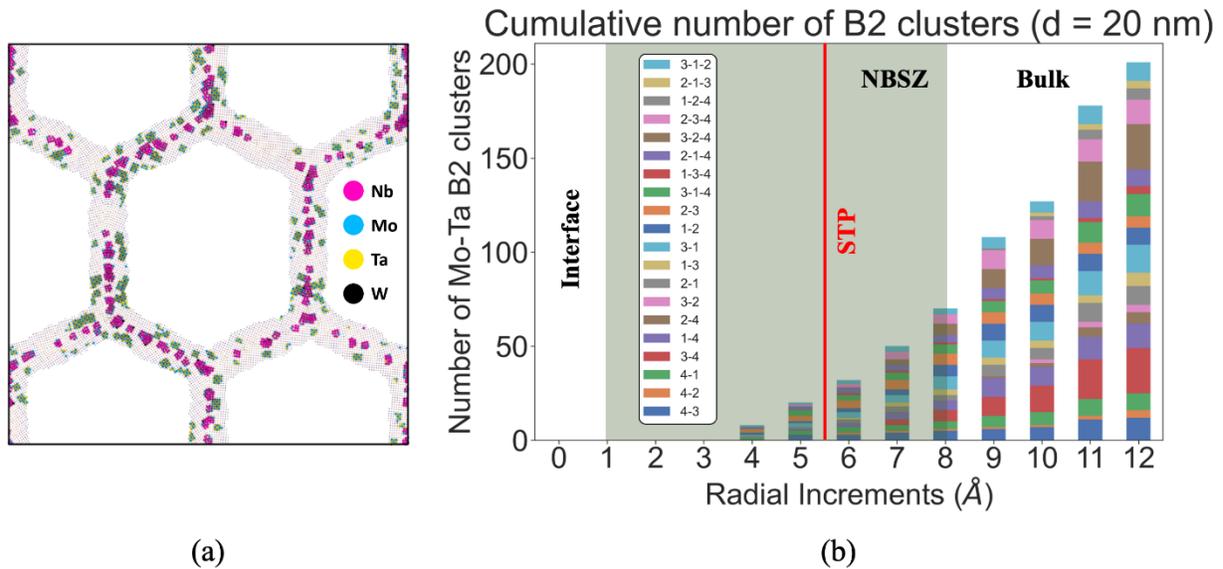

(a)                                                        (b)

**FIG. 10. (a) 20 nm grain size model showing that Nb-Nb clusters with A2 structures (pink) dominate the interface region while the Mo-Ta B2 structures (yellow and blue) are primarily in the NBSZ. (b) Cumulative Mo-Ta B2 cluster counts for the grain boundary and triple junction regions of the 20 nm grain size model at 300 K, showing significant boundary-to-boundary variation.**

In order to investigate the structural transitioning in individual regions, the same slicing technique applied to all interface sites is also applied to obtain the behavior at individual grain boundaries. We note that this type of radial slicing technique applied to triple junctions results in overlapping sites at the intersection of triple junctions and grain boundaries, which appear purple in Fig. 2(b). In other words, some sites a certain distance away from a triple junction would actually be located at the core of a grain boundary. This is also the reason for selecting the largest grain size model instead of the smaller grain sizes for the interfacial CSRO analysis (i.e., to utilize more data points belonging to grain boundary regions). Therefore, we limit our analysis to the planar grain boundaries of the 20 nm grain size model alone here. Supplementary Note 4 shows interface atomic fractions and atomic volumes associated with grain boundary regions that are used in the calculations of NBSZs and STPs of each grain boundary. Nb-Nb and Mo-Ta chemical order parameters and their corresponding NBSZs and STPs are shown for the grain boundary regions in Fig. 11. Grain boundaries are ordered from top-to-bottom and left-to-right in decreasing number of vicinal Mo-Ta B2 clusters, and all grain boundary regions exhibit the antagonistic A2-to-B2 transition. The variation in NBSZ and STP width at individual grain boundary regions stems from the interfacial character.[22]

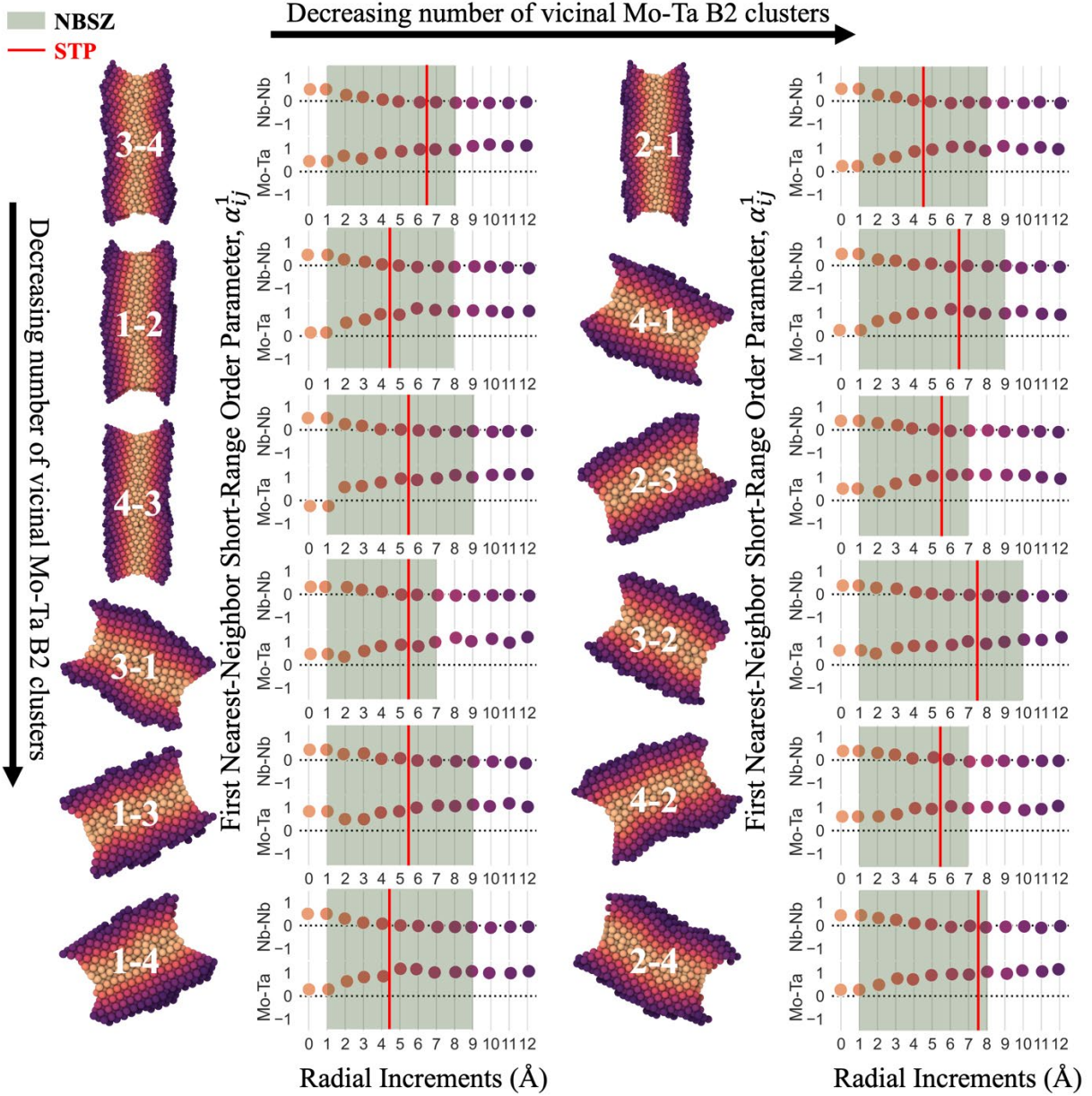

**FIG. 11. Chemical short-range order behavior of Mo-Ta and Nb-Nb pairs for individual grain boundary regions. Grain boundaries are ordered by their number of vicinal Mo-Ta B2 clusters.**

# 4   CONCLUSIONS

In this work, the effect of interfacial structure on chemical ordering is investigated in a refractory NbMoTaW CCA.  Hexagonal grain models with three different grain sizes (5, 10 and 20 nm) are created and hybrid MC/MD simulations are performed at temperatures ranging from 100 to 2000 K to obtain a wide variety of interfacial segregation patterns. Interface regions are sliced and partitioned to investigate near-boundary segregation and structural transitions within the overall boundary network, as well as individual grain boundary and triple junction regions. The crystallographic orientations considered in this work should be generic in nature, as the tilt angles are arbitrary and the boundary structures very random (i.e., no repeating units or special orientations). Some important observations from this study are:

(1) At the interface regions, heavy Nb enrichment and proportionate W and Ta depletion are observed. The effect of the grain size on segregation behavior at the interfaces seems marginal for these three models.

(2) Investigating the segregation behavior in radially expanded slices from interfaces to bulk reveals that the interfacial defects alter composition in regions that span much farther than just the defected atoms. These NBSZs can make the boundary-affected region 2-3 times larger than the width associated with defect atoms alone.

(3) Within these NBSZs, the predominant A2 structure of the interface (Nb-Nb) is taken over by the B2 structure (Mo-Ta) at the STP as one moves away from the grain boundary core.

(4) These STPs were not identified in the prior study by McCarthy et al. [50] for a family of FCC CCAs, indicating the importance of the CSRO and clustering behavior in this alloy.

(5) The NBSZ and the STP widths are both dependent on grain size and temperature. These widths decrease as temperature increases up to the order-disorder transition temperature, after which they become invariant to further temperature increases.

(6) Analysis of the CSRO suggests that not all interactions are affected by the presence of an interface. For instance, the Nb-Ta pair exhibits the same ordering tendencies regardless of its location.  In contrast, the Mo-Ta pair has the strongest ordering tendency in the bulk region but much weaker ordering tendency at the interfaces.

(7) The two most frequent cluster types observed are the A2 (Nb-Nb) and B2 structures (Mo-Ta).  Chemical clustering analysis on the radially expanded slices indicates that Nb-Nb and Mo-Ta exhibit antagonistic CSRO behavior.  Starting from interfaces and moving into the bulk region, Nb-Nb transitions from ordered-to-disordered whereas Mo-Ta transitions from disordered-to-ordered.

(8) Analysis on individual regions suggests that some grain boundaries are more inclined to harbor clusters.  All grain boundary regions exhibit the A2-to-B2 transition, albeit at different atomic layers away from interface, which supports the hypothesis that the interface structure alters chemical ordering and clustering tendencies.

As a whole, the findings of this study demonstrate how different parameters such as grain size, temperature, and crystallographic orientation modify segregation both at and near the interfacial regions in RCCAs with strong CSRO.  These findings can facilitate microstructural design of CCAs through altering the properties affecting the NBSZ and STP widths; such as grain size, temperature, and interface character.  Then, this engineered property could have practical applications in improving complex mechanisms such as dislocation pinning, solid solution hardening, electromagnetism, and radiation damage healing.

## 5    SUPPLEMENTARY MATERIAL

See supplementary material for additional details on the convergence of the gradient of potential energy during the hybrid MC/MD procedure, cut-off lengths, details pertaining to slicing and partitioning techniques, and grain boundary region plots.

# 6    DATA AVAILABILITY STATEMENT

The data that supports the findings of this study are available within the article and its supplementary material.

# 7    ACKNOWLEDGEMENTS

This research was primarily supported by the National Science Foundation Materials Research Science and Engineering Center program through the UC Irvine Center for Complex and Active Materials (DMR-2011967). The authors thank Prof. Shyue Ping Ong for the helpful discussions regarding machine learning interatomic potential utilized in this work. This article has been authored by an employee of National Technology & Engineering Solutions of Sandia, LLC under Contract No. DE-NA0003525 with the U.S. Department of Energy (DOE). The employee owns all right, title and interest in and to the article and is solely responsible for its contents. The United States Government retains and the publisher, by accepting the article for publication, acknowledges that the United States Government retains a non-exclusive, paid-up, irrevocable, world-wide license to publish or reproduce the published form of this article or allow others to do so, for United States Government purposes. The DOE will provide public access to these results of federally sponsored research in accordance with the DOE Public Access Plan https://www.energy.gov/downloads/doe-public-access-plan.